\documentstyle[osa,twocolumn]{revtex}

\begin{document}
\title{Quantum control by compensation of quantum fluctuations}

\author{Holger F. Hofmann and Ortwin Hess}
\address{Institute of Technical Physics, DLR,
       Pfaffenwaldring~38-40, D--70569 Stuttgart, Germany}
\author{G\"unter Mahler}
\address{Institute of Theoretical Physics and Synergetics, 
       Pfaffenwaldring~57, D--70550 Stuttgart, Germany}
\maketitle

\begin{abstract}
We show that the influence of quantum fluctuations in the electromagnetic 
field
vacuum on a two level atom can be measured and consequently compensated by 
balanced homodyne detection and a coherent feedback field. This compensation
suppresses the decoherence associated with spontaneous emissions for a 
specific state of the atomic system allowing complete control of the 
coherent state
of the system.       
\end{abstract}

\vspace{1cm}
Attempts to control the states of quantum systems often provide new insights
into the fundamental nature of quantum mechanics and reveal new aspects of the
transition from classical to quantum mechanical behaviour. 
The reason for this is that the concept of quantum control requires us to 
examine details of the effects causing decoherence which may have been 
overlooked before. One typical effect causing decoherence is the interaction 
of excited atoms with the electromagnetic vacuum giving rise to spontaneous
emission. It is especially important
since coherence is often established by electromagnetic fields, requiring 
the quantum system to be open to a continuum of modes. The conventional way
of dealing with the problem of decoherence in the presence of spontaneous 
emission is to distinguish no-photon intervals and photon emission events
\cite{Ple97,Gar91,Car93}. However, this is by no means
the only way of 
observing the electromagnetic field propagating away from a quantum system.
As pointed out by Ueda \cite{Ued92}, a measurement of the emitted field which
is sensitive to the vacuum state as well is logically reversible, as opposed
to the sudden transition to the ground state connected with a photon detection
event. Therefore it seems preferable to apply measurement schemes different 
from photon detection if quantum coherence is to be controlled.

In this letter we consider the possibility of observing one quadrature
component of the  
electromagnetic field propagating away from the atomic system by time 
resolved balanced homodyne detection.
The field actually originating from the dipole oscillations of the 
atomic system on a timescale $\tau$ which is much smaller than the 
lifetime $1/\Gamma$ of the 
excited atomic state is much smaller than the vacuum fluctuations observed on 
this timescale.
Thus it is possible to interpret the fields measured as quantum fluctuations 
of the electromagnetic field impinging on the system. In this sense the 
measurement is a measurement of the forces acting on the system and not a
measurement of the system state itself. 
It should be possible to compensate the effect of the observed quadrature
component of the electromagnetic field by a coherent field of opposite sign. 
However, the effect of the unobserved
quadrature component must also be compensated if decoherence is to be 
suppressed. 
To find out, how this can be achieved as well, it
is necessary to investigate the back-action of 
the homodyne detection on the atomic system.

For the description of the homodyne detection process, we use a 
non-orthogonal projective measurement base. 
This type of measurement base for homodyne detection has been 
derived and applied in a number of publications
\cite{Yue78,Bra90,Vog93,Lui96}. 
Since the observed fields are small, we will only consider that
part of the measurement base composed of the zero or one photon contributions.
The effective non-orthogonal measurement base is given by                 
\begin{eqnarray}
\label{eq:project}
\mid \mbox{P}(\Delta n)\rangle  &=& (2\pi\alpha^*\alpha)^{-1/4} 
\exp[-\frac{\Delta n^2}{4\alpha^*\alpha}]
\nonumber \\
& & \times
\left(\mid \mbox{vacuum}\rangle  + \frac{\Delta n}{\alpha^*}\mid n_{in}=1\rangle \right),
\end{eqnarray}
where $\alpha$ is the field amplitude of the coherent field mode emitted by the
local oscillator during the time segment $\tau$ considered in the measurement 
and $\Delta n$ is the measured photon number difference between the two 
branches of the homodyne detection setup. Note that $\alpha$ is related to the
intensity (or photon rate) $I$ emitted by the local oscillator by 
$I=\alpha^*\alpha/\tau$. The relation has been derived using the assumption 
that $\alpha^*\alpha \gg 1$. Details of the derivation will be given elsewhere
\cite{Hof97}.
Within the zero- and one-photon subspace weak coherent fields of amplitude $\beta$
are  approximately given by
\begin{equation}
\mid \Phi_{\beta}\rangle  \approx \mid \mbox{vacuum}\rangle  + \beta \mid n_{in}=1\rangle .
\end{equation}
The measurement probability of a photon number difference $\delta n$ of such a coherent field
can be calculated from equation (\ref{eq:project}) by
\[
p_{\beta}(\Delta n) 
= \mid \langle \mbox{P}(\Delta n)\mid \Phi_{\beta}\rangle \mid^2
\]
\begin{equation}\label{eq:beta}
= \frac{1}{\sqrt{2\pi\alpha^*\alpha}} 
  \exp[-\frac{(\Delta n-(\alpha^*\beta+\beta^*\alpha))^2}{2\alpha^*\alpha}].
\end{equation}
This is a Gaussian with a variance of 
$\langle \Delta n^2\rangle = \alpha^*\alpha$ and a mean value of 
$\langle \Delta n \rangle = \alpha^*\beta+\beta^*\alpha$. If the measured
value of $\Delta n$ is identified as $2 \mid \alpha \mid$ times the quadrature
component of the measured light field in phase with the local oscillator,
this result exactly corresponds to the quantum uncertainty of $1/4$ and a
shift by the component of $\beta$ in phase with $\alpha$. This result confirms 
the interpretation of homodyne detection as a projective measurement of the
quadrature component in phase with the local oscillator.

The dynamics of the photon emission process and the interaction of a two
level atom with the light field continuum can be analyzed without assuming
an optical cavity or using a bath approximation by applying Wigner-Weisskopf
theory to the complete system-field Hamiltonian \cite{Hof95}. 
In the following,
however, we will assume fast time-resolved measurements performed on the 
field long before the emission probability from an excited state approaches unity. 
During the short time intervals $\tau$ with 
$\Gamma\tau \ll 1$, the one-photon component of the wavefunction corresponds to
a photon in a field mode with a rectangular envelope: zero field amplitude for
distances $r$ from the atomic system with $r>c\tau$ and a constant probability
of finding a photon at distances of $0<r<c\tau$. Therefore, the photon 
possibly 
emitted during the time interval $\tau$ is in a well defined mode. Thus it is
possible to write down the wave function which evolves from the light field 
vacuum and an arbitrary state of the two level atom given by
\begin{equation}
\label{eq:psi0}
\mid \Psi(0)\rangle  = c_E \mid \tilde{E};\mbox{vacuum}\rangle  
                     + c_G \mid G;\mbox{vacuum}\rangle, 
\end{equation}
where $\mid G \rangle$ is the atomic ground state and $\mid \tilde{E}\rangle$
is the excited state in the interaction picture, i.e.~without the phase dynamics 
at the frequency $\omega_0$ of the atomic transition. 
After the time interval $\tau$, the entangled state of the atomic system and 
the electromagnetic field is
\[
\mid \Psi(\tau)\rangle  = 
                 c_E (1-\Gamma\tau/2)\mid \tilde{E};\mbox{vacuum}\rangle  
\]
\begin{equation}
\label{eq:psitau}
                        + c_G \mid G;\mbox{vacuum}\rangle 
                        + c_E \sqrt{\Gamma\tau}\mid G;n_0=1\rangle .
\end{equation}
This is the complete quantum mechanical state as it evolves unitarily according to
the Hamiltonian of Wigner-Weisskopf theory. 

The projective measurement base given in equation (\ref{eq:project}) may now be applied to determine
the change in the state of the atomic system conditioned by a measurement
of $\Delta n$ in the homodyne detection during the time interval $\tau$.
The wavefunction $\mid \psi(\tau)\rangle$ of the atomic system after the measurement reads
\begin{eqnarray}
\mid \psi(\tau)\rangle  
& = & \langle \mbox{P}(\Delta n) \mid \Psi (\tau)\rangle  \nonumber\\ 
&& \nonumber \\ 
& = & (2\pi\alpha^*\alpha)^{-1/4}\exp[-\frac{\Delta n^2}{4\alpha^*\alpha}]
\nonumber 
\end{eqnarray}
\begin{equation}
\times 
\left(c_E (1-\Gamma\tau/2)\mid \tilde{E}\rangle 
+(c_G + c_E \sqrt{\Gamma\tau}\frac{\Delta n}{\alpha})\mid G\rangle \right).
\end{equation}
Since $\Gamma\tau \ll 1$, the squared length of this state vector which 
corresponds to the probability of measuring $\Delta n$ is approximately
independent of the system state and is given by the vacuum distribution,
\begin{equation}
\label{eq:prop}
p(\Delta n)
    \approx \frac{1}{\sqrt{2\pi\alpha^*\alpha}} 
             \exp[-\frac{\Delta n^2}{2\alpha^*\alpha}].
\end{equation}
The major contribution to the change of the state of the atomic system 
conditioned by the measurement is given by the amplitude proportional to
$\sqrt{\Gamma\tau}$. The higher order terms do have some effect on 
timescales of $1/\Gamma$, corresponding to a large number of measurement 
intervals $\tau$. These effects will be discussed elsewhere \cite{Hof97}. 
In the following
we will concentrate on the short time fluctuations effective on a timescale
of $\tau$. 

If the normalized system state is written as
$\mid \psi(0)\rangle + \mid \delta \psi (\tau)\rangle$, such that 
$\mid \delta \psi (\tau)\rangle$ is the change of the system state orthogonal 
to $\mid \psi(0)\rangle$, then this change is approximately given by
\begin{eqnarray}
\label{eq:svdiff}
\mid \delta\psi(\tau)\rangle 
                   &\approx& -\sqrt{\Gamma\tau}\frac{\Delta n}{\mid\alpha\mid}
c_E^2\left(c_G^* \mid \tilde{E}\rangle  - c_E^* \mid G\rangle \right).
\end{eqnarray}
Since the probability distribution of measurement results  $\Delta n$ is a Gaussian, this
equation describes a diffusion process. Statistically, the diffusion steps
cancel on average, causing decoherence because the uncertainty of the actual
path chosen by the system dynamics increases with each unknown step. 
In our scenario however, the length and the direction of each step has been
measured by homodyne detection. 

We can therefore deduce the evolution of
the pure state of the atomic system. 
In this sense the description of the
quantum measurement process is a quantum trajectory description as introduced
in \cite{Car93,Car96} and applied to problems of continuous feedback 
scenarios 
in \cite{Wis94}. It has not been derived from a master equation of the open
system, however, and the field-atom interaction is described using the 
Schroedingers equation of Wigner-Weisskopf theory, retaining the full 
atom-field entanglement up to the projective measurement. 

In order to visualize the diffusion step, it is useful to describe the 
state of the atomic system by its Bloch vector ${\bf s}$, defined as
\begin{eqnarray}
s_{x} &=& 2 \mbox{Re}\left(\langle \psi\mid \tilde{E}\rangle \langle G\mid \psi\rangle \right) \\
s_{y} &=& 2 \mbox{Im}\left(\langle \psi\mid \tilde{E}\rangle \langle G\mid \psi\rangle \right) \\
s_{z} &=& \mid \langle \tilde{E}\mid \psi\rangle \mid^2 - \mid \langle G\mid \psi\rangle \mid^2,
\end{eqnarray}
where $\mbox{Re}(\cdot)$ and $\mbox{Im}(\cdot)$ denote the real and imaginary part, respectively.
$s_z$ is the expectation value of the population inversion and $s_x$
and $s_y$ are the in-phase and the out-of-phase dipole moments 
of the atomic system, respectively. The change in the Bloch vector of the atomic system
$\delta{\bf s}$
conditioned by a measurement of $\Delta n$ within the time interval 
$\tau$ is then given by
\begin{equation}
\left(\begin{array}{c}\delta s_{x}\\ \delta s_{y}\\ \delta s_{z}\end{array}
\right) =  \sqrt{\Gamma\tau}\frac{\Delta n}{\mid\alpha\mid}
\left(\begin{array}{c}s_z+1-s_x^2\\-s_xs_y\\-s_x-s_xs_z\end{array}
\right). 
\end{equation}
A representation of this diffusion step on the Bloch sphere is shown in 
Fig.~1. 

The linear part of this change in the Bloch vector corresponds to a 
Rabi rotation around $s_y$. It is exactly equal to the effects of a
coherent field with an amplitude of $\Delta n/\mid 2 \alpha \mid$.
The non-linear part is shown in Fig.~2.

For positive $\Delta n$, this contribution draws the Bloch vector towards the 
$s_x = +1$ pole of the Bloch sphere. For negative $\Delta n$, the
Bloch vector moves towards the $s_x=-1$ pole. 

It is possible to interpret this effect of the quantum 
fluctuations on the atomic system as an epistemological effect of 
information on the 
in-phase dipole component $s_x$ gained in the measurement. Positive
values of $\Delta n$ make a positive dipole component $s_x$ more likely
and negative values of $\Delta n$ make a negative dipole component $s_x$ 
more likely. Although the information obtained in a single measurement is
almost negligible, the relative suppression of the amplitude of one
dipole eigenstate and the corresponding amplification of the amplitude
corresponding to the other dipole eigenstate causes a change in the 
state of the atomic system unless the system is already in an eigenstate
of the dipole component with $s_x=\pm 1$. The relative smallness of the 
dipole field compared to the quantum fluctuations makes this measurement
a weak measurement in the sense discussed by Aharonov and coworkers in
\cite{Aha88}. 

Even though the non-linear dependence of the diffusion step on the 
previous state of the atomic system prevents a state independent compensation
of quantum fluctuations, the measurement is still logically reversible in
the sense of \cite{Ued92}. It can be compensated if the previous state of
the system is known with sufficient precision. 
In the following, we shall focus on atomic system states with $s_y=0$. For such states, 
$\delta s_y$ is also zero and the whole diffusion process takes place
in the $s_x,s_z$ plane. The diffusion steps may then be identified as 
rotations around the $s_y$ axis. By defining the angle $\theta$ such that
$\cos \theta = s_z$ and $\sin \theta = s_x$, the diffusion step in the 
$s_x,s_z$ plane may be written as
\begin{equation}
\delta\theta = \sqrt{\Gamma\tau}\frac{\Delta n}{\mid\alpha\mid}(1+\cos \theta).
\end{equation}
This rotation of the Bloch vector conditioned by the measurement of $\Delta n$
is equivalent to a Rabi rotation around the $s_y$ axis proportional to the
quadrature component
measured in the homodyne detection. Despite the quantum mechanical dependence
of this Rabi rotation on $\theta$, it is possible to compensate 
the effects of the quantum fluctuations by simply reversing the rotations
corresponding to each measurement. The feedback field $f$ necessary to 
stabilize a state of the atomic system with $\theta = \bar{\theta}$ is given by
\begin{equation}
f(\Delta n_{0})=-(1+\cos\bar{\theta})\frac{\Delta n_{0}}{2 \mid \alpha \mid},
\end{equation}
where $\Delta n_{0}$ is the measurement result associated with the quantum 
fluctuations which are to be compensated by the feedback term. Each time
interval $\tau$ is therefore associated with a diffusion step caused by
the quantum fluctuations and a time delayed feedback which compensates the
diffusion step. 
The total field interacting with the atomic system 
is given by a coherent state of field amplitude $f(\Delta n_{0})$. This 
corresponds to vacuum-state quantum fluctuations shifted by $f(\Delta n_{0})$.
Consequently, the measurement result
$\Delta n_{next}$ corresponding to the time interval $\tau_{next}$ during which
the feedback field acts on the system will be composed of a stochastic
effect of the quantum fluctuations $\Delta n_{qf}$ and a shift $\delta_{next}$
caused by the feedback field
\begin{eqnarray}
\Delta n_{next} &=& \Delta n_{qf} + \delta_{next} \nonumber \\
                &=& \Delta n_{qf} + 2 \mid \alpha \mid f(\Delta n_0).
\end{eqnarray}
Since the feedback effect itself should not be compensated, only the 
contribution of the quantum fluctuations $\Delta n_{qf}$ should be 
applied for the determination of the subsequent feedback field. 

Effectively, the atomic system now interacts with a series of light field
modes initially in weak coherent states corresponding to vacuum 
fluctuations shifted by the feedback field. The 
quadrature component of
the field in phase with the local oscillator is measured, revealing the
amplitude of this component of the fluctuations. The feedback then correlates
the average field of the next light field mode interacting with the system
with the measured quadrature component of the fluctuations. In a classical
system this compensation would be insufficient since the out-of-phase
quadrature component of the fluctuating field is unknown. In the quantum
mechanical case the information obtained is complete. 
Instead of causing uncontrollable changes in the system state, the effect
of the quantum fluctuations corresponds to a weak measurement correlating the
information gained in the field measurement with the information about the
system state. This effect is therefore predictable and can be compensated.
The sum of the measured fluctuations and the feedback field reveals which
part of the feedback field is necessary to compensate the changes associated 
with the weak measurement of $s_x$:
\begin{equation}
\frac{\Delta n}{2 \mid \alpha \mid} + f(\Delta n) = 
-\cos\bar{\theta}  \frac{\Delta n}{2 \mid \alpha \mid}.
\end{equation}
The state dependence of the weak measurement can be illustrated 
in terms of the three most typical cases:

{\em Dipole eigenstate}. 
For $\cos \bar{\theta} = 0$ the system is in an eigenstate of the in-phase 
dipole component $s_x$. No measurements of $s_x$, whether weak or strong, 
will change this. Therefore, the compensating field necessary to suppress
the effects of quantum fluctuations is equal to the compensation of 
the classically expected Rabi rotation. Also note that a coherent field 
along the unknown field quadrature would not affect this state, since 
the Bloch vector is parallel to the axis of Rabi rotations caused by 
fields $\pm\pi/2$ out of phase with the local oscillator. 

{\em Ground state}. For $\cos \bar{\theta} = -1$ no feedback is necessary
for stabilization. This means that the effects of the Rabi rotation and
the weak measurement associated with a homodyne detection result $\Delta
n$ automatically compensate each other. This is a result of the fact that
the ground state is polarized by the field in such a way that the dipole
emissions interfere destructively with the field. At the same time, 
the observed field makes a dipole more likely which emits radiation 
interfering constructively with the fluctuations. This effect may also be
understood in terms of energy conservation. The ground state atom absorbs
the field by the destructive interference of dipole emission and incoming
field, but at the same time it emits radiation associated with the
quantum fluctuations of the dipole variables. Both effects cancel and energy
conservation is preserved. 
  
{\em Excited state}. For $\cos \bar{\theta} = +1$, the feedback necessary to
compensate the weak measurement effects is equal to the feedback necessary
to compensate the Rabi rotations. The reason for this is that the excited
state is polarized by the field in such a way that the dipole emissions 
interfere constructively with the field. At the same time, the measurement
makes such a dipole more likely. Consequently the effect of the 
quantum fluctuations is doubled. In terms of energy conservation the excited
state atom amplifies the field and emits additional radiation associated
with the quantum fluctuations of the dipole. The instability of the excited 
state is thus related to its linear response to the light field which implies
gain instead of absorption. The feedback field corrects this property by
effectively reversing the sign of the susceptibility, overcompensating the
loss in energy associated with the field amplification and establishing a
stability equivalent to that of the ground state without feedback.

If the quantum state of the system does not correspond to the state for
which the diffusion step is suppressed by the feedback, the total effect
of quantum fluctuations and a feedback signal given by $\cos \bar{\Theta}$
may be expressed by a general diffusion step on the Bloch sphere. 
This diffusion step reads
\begin{equation}
\left(\begin{array}{c}s_x 
\\ s_y     \\ s_z \end{array} \right) = \sqrt{\Gamma\tau}\frac{\Delta n}{\mid \alpha \mid}
\left(\begin{array}{c}-\cos\bar{\Theta}s_z+1-s_x^2 
\\ -s_xs_y \\ + \cos\bar{\Theta}s_x - s_xs_z 
\end{array} \right).
\end{equation}
Note that the diffusion step for any state with $s_z = \cos\bar{\Theta}$
is perpendicular to the $s_z$ axis, indicating that the feedback stabilizes 
this value of the inversion expectation value $s_z$ regardless of the 
phase of the dipole oscillations relative to the local oscillator.

In conclusion, we have shown that homodyne detection of the electromagnetic
field propagating from a single two level atom with a known initial
quantum state reveals the changes induced in the state of the atom by  
quantum fluctuations.
In the $s_x,s_z$ plane of the Bloch vector representation of the
atomic system, randomly fluctuating forces cause rotations around
the $s_y$ axis. This effect corresponds to that of a coherent driving field
and can consequently be compensated by Rabi rotations of opposite sign
induced by a feedback field.  
The decoherence caused by quantum fluctuations
can be suppressed completely with a precision limited only by the time delay
between the emission of the field and the measurement by homodyne detection.
Even though one quadrature component of the light field remains unobserved,
we have demonstrated that quantum control of an arbitrary state of a two-level atomic system 
is possible by simply applying a coherent feedback field. 


%
\begin{figure}
\caption{Visualization of the diffusion step on the Bloch sphere. 
The diffusion
is represented by lines oriented parallel to the direction of the diffusion
with a length proportional to the standard deviation of the step length.
a) and b) show the projections into the $s_y, s_z$ and the $s_x,s_z$ plane, respectively.}
\end{figure}

\begin{figure}
\caption{Non-linear contribution to the diffusion step of the Bloch vector.
The representation is in analogy to Fig.~1, with~a) and b) showing the projections into the $s_y, s_z$ and into 
the $s_x,s_z$ planes, respectively.}
\end{figure}
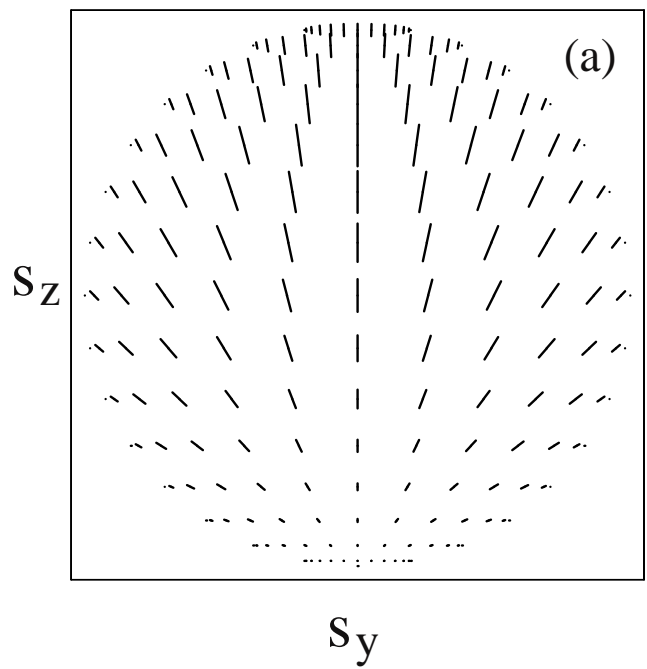
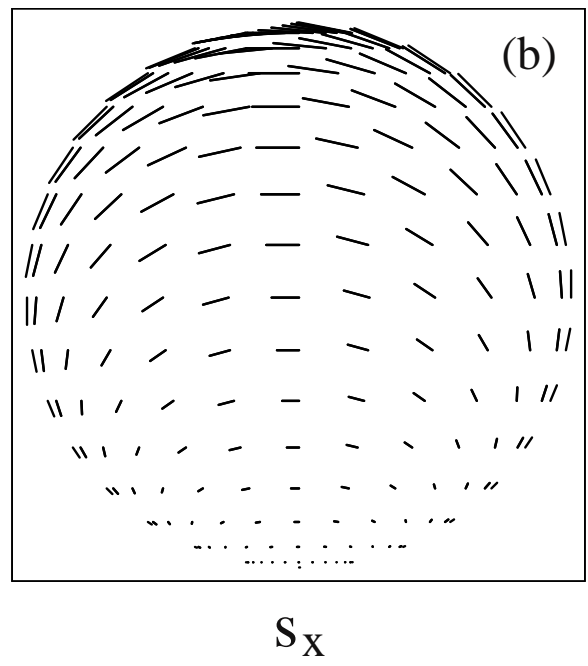
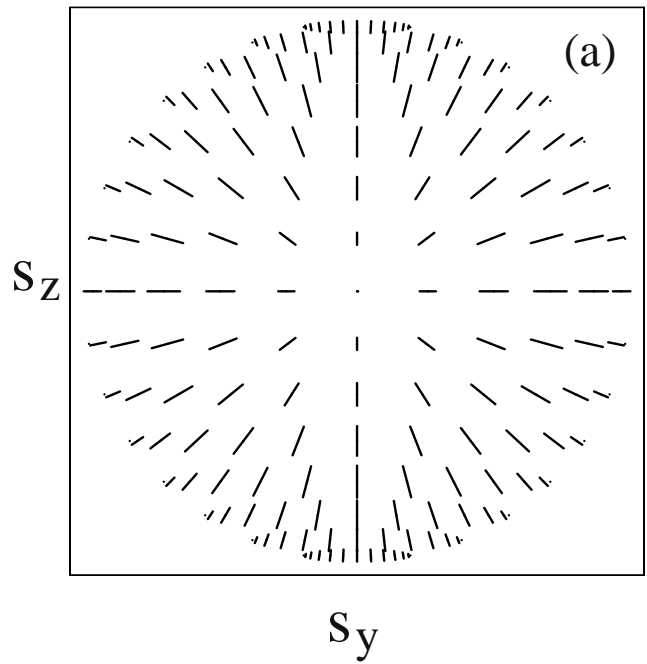
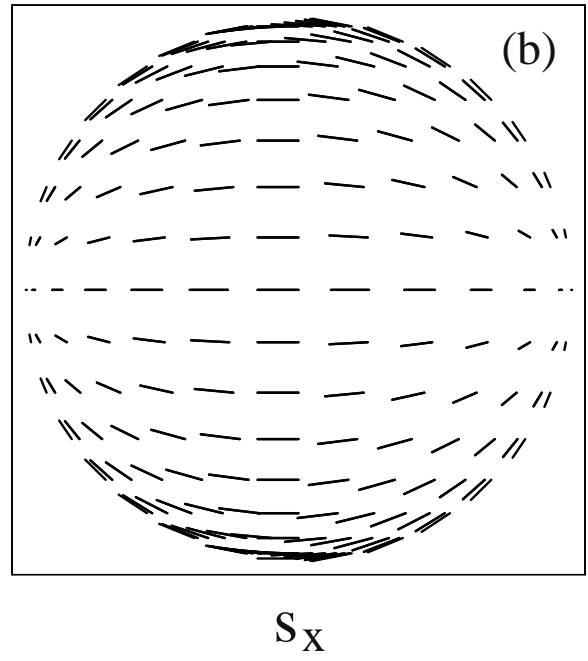
\end{document}